\documentstyle{ichep}
\begin{document}

\title{Gluon Radiation Patterns in Pomeron Exchange Events
\footnote{Talk given at the {\it XXVII International Conference on High
Energy Physics}, Glasgow, Scotland, July 21--27, 1994.  }}

\author{Dieter Zeppenfeld }

\affil{Department of Physics, University of Wisconsin,
 Madison, WI 53706, USA }

\abstract{Color singlet two gluon exchange provides a perturbative model for
the pomeron. This mechanism is thought to explain the production of rapidity
gaps in hard dijet events at the Tevatron. It is shown that in $qQ$ scattering
via two gluon color singlet exchange the emission of soft gluons follows
closely the pattern found for $t$-channel photon exchange. Gluon emission is
strongly suppressed between the two quark jets. After hadronization this leads
to a depressed level of hadronic activity between the jets and thus allows the
formation of rapidity gaps.  }

\twocolumn[\maketitle]

\fnm{1}{E-mail: dieter@pheno.physics.wisc.edu}

%\section{}
The formation of rapidity gaps, regions in pseudorapidity without hadronic
activity, is a well known phenomenon in hadronic collisions. Elastic
scattering or single diffractive dissociation %in {\it e.g.} $pp$ collisions
are examples of low $Q^2$ processes with rapidity gaps and they have
long been understood in terms of pomeron exchange~\cite{Grib,lan}.

A similar phenomenon has recently been observed
in {\it hard} scattering events at the Tevatron~\cite{brandt}. The D0
Collaboration has studied a sample of dijet events with jet transverse energies
in excess of 30 GeV. In about $ 0.5$\%  of all events with widely
separated jets no sign of hadronic activity is observed  between the two jets.
Sampling hadrons between the two jets, a clear break at low multiplicities
is seen in the multiplicity distribution. This signals the existence
of a qualitatively new source of forward parton scattering. Such events
have been predicted~\cite{bjgap}, at the observed level, in terms of the
$t$-channel exchange of two gluons in a color singlet state which is the
Low-Nussinov model for the pomeron~\cite{low}.

The exchange of a $t$-channel color singlet object, like the pomeron or a
photon, can be seen to lead to rapidity gaps in the color string picture.
In such events color is restored by forming a string between the forward
scattered partons and their respective beam remnants. Hence, hadrons are
predominantly produced in the forward and backward
regions, leaving a rapidity gap between the two scattered partons.

Alternatively, the distribution of hadrons can be understood in terms of the
pattern of (typically  soft) radiated gluons in the hard scattering event. In
the case of $\gamma$ exchange in forward $qq \rightarrow qq$ scattering, color
coherence~\cite{colcoh} between initial and final state gluon radiation
is known to lead to an exponentially suppressed gluon emission probability
into the rapidity region between the two final state quarks~\cite{fletcher}.
This suppression then leads to the formation of rapidity gaps. The marked
difference in the gluon radiation patterns of $t$-channel photon vs.
$t$-channel gluon exchange is demonstrated in Fig.~\ref{figone}a.
\begin{figure*}
\vspace*{8.2cm}
\caption{Rapidity distribution of emitted gluons in $qQ\to qQg$ scattering for
fixed final state parton transverse momenta of $p_{Tq}=30$~GeV and
$p_{Tg} = 2$~GeV. The quark rapidities are fixed at $\eta_q = \pm 3$
(indicated by the arrows). In part a) results are shown for the sum over all
color structures for single gluon and for
$t$-channel photon exchange.   The $M_{34}$ terms alone, in part b),
demonstrate
the difference between the QED and the QCD color singlet exchange terms.
\label{figone}}
\end{figure*}
Due to its
color singlet structure one expects that the same would occur for the
$t$-channel exchange of a pomeron. However, since the Low-Nussinov pomeron is
an extended object with colored constituents, the validity of this analogy
is not obvious: gluon radiation may resolve the internal color structure of
the pomeron. This talk discusses a recent analysis of this problem~\cite{CZ}.

It is instructive to first study the process
$q_{i_1}Q_{i_3}\to q_{i_2}Q_{i_4}g^a$, mediated via photon
(QED) or single gluon  exchange (QCD), at the tree level.
The general scattering amplitude for this process can be decomposed into
two color singlet and two color octet exchange amplitudes. Here, we only need
to consider color singlet exchange as viewed by quark $Q$ which is given by the
$M_{34}$ coefficient in the color decomposition of the amplitude,
\begin{equation}\label{M34def}
M = {\frac {\lambda_{i_2i_1}^{a}} {2} }\; {\delta_{i_4i_3}}\; M_{34} + ... \, .
\end{equation}
Even for $t$-channel gluon exchange this color singlet amplitude exists.
However, the rapidity distribution of the emitted gluon is markedly different
from photon exchange. In the QED case the
$M_{34}$ amplitude corresponds to emission of the final state gluon off the
quark $q$. In forward scattering ($\eta_q=+3$ in Fig.~\ref{figone}) the gluon
is radiated between the initial and final state $q$ directions.
The color $i_1$ of the initial quark $q$ is
thus transferred to a low mass color triplet object which emerges close to
the beam direction. At lowest order this is the final state $q$, at
${\cal O}(\alpha_s)$ it is the $qg$ system. The situation is thus stable
against gluon emission at even higher order for the QED case and gluon
radiation is suppressed in the rapidity range between the two final state
quarks.

In the QCD case $M_{34}$ corresponds to emission of the gluon from the
quark $Q$.
%the color transferred to $Q$ by the $t$-channel gluon needs to be carried
%off again by the radiated gluon.
The gluon is preferentially emitted between the initial $Q$-beam and the
final $Q$ directions (dash-dotted line in Fig.~\ref{figone}b).
Thus the color triplet $qg$ system, into which the
initial quark $q$ evolves, consists of a widely separated quark and gluon.
Higher order corrections will lead to strong gluon radiation into the angular
region between the two and thus also into the rapidity range between the
two final state quarks.

These typical patterns found for $t$-channel color singlet and color octet
exchange may now be used as a gauge for the radiation pattern produced in
$qQ\to qQg$ scattering via the exchange of two gluons in a color singlet state.
In the lowest order process, $qQ\to qQ$, the color singlet exchange amplitude
is dominated by its imaginary part~\cite{cudell}. Hence, we may estimate the
radiation pattern
by calculating the imaginary part of the gluon emission amplitude $M_{34}$
only. Typical Feynman graphs are shown in Fig.~\ref{figtwo}. Details of the
calculation are given in Ref.~\cite{CZ}.
\begin{figure}
\vspace*{4.0cm}
\caption{Two of the 31 Feynman graphs contributing to the imaginary part of
the color singlet exchange amplitude $M_{34}$.    \label{figtwo}}
\end{figure}
\begin{figure*}
\vspace*{8.2cm}
\caption{Rapidity distribution of emitted gluons in $uc\to ucg$ scattering
via two gluon color singlet exchange as seen by the charm quark.
%The solid line gives the result for pomeron exchange and a regularizing
%gluon mass parameter of $m_r=300$~MeV.
The phase space
parameters for the quarks are the same as in Fig.~1 and results are shown
for a) the case of a soft gluon ($p_{Tg}=2$~GeV) and b) a hard gluon
($p_{Tg}=15$~GeV). For comparison tree level results are shown for gluon
(dash-dotted lines) and photon exchange (dashed lines).
 \label{figthree}}
\end{figure*}

For massless internal gluon propagators the phase space integrals over the
$qQ$, $qg$, and $gQ$ intermediate states are divergent. They can be
regularized by replacing the massless gluon propagator by a version which
avoids unphysical gluon propagation over long distances~\cite{lan}. QCD
Pomeron models of this kind have been found to give a good description of
available data~\cite{natale}. These refinements can be approximated by using
an effective gluon mass of $m_r = 300$~MeV in the calculation.

A second problem arises because some of the contributions to ${\rm Im}M_{34}$
correspond to $q\to g$ splitting and subsequent $gQ\to gQ$ scattering via
pomeron exchange. These contributions cannot be expected to be suppressed
when the gluon is emitted between the $q$ and the $Q$ directions and thus
would mask the radiation off pomeron exchange in $qQ$ scattering. These
splitting contributions have been subtracted in Ref.~\cite{CZ} to yield the
square of the pomeron exchange radiation pattern,
$|{\rm Im}M_{34}^{\rm pom}|^2$, which is shown in Fig.~\ref{figthree}.

For high transverse momentum of the emitted gluon (of order of the quark
momenta, see Fig.~\ref{figthree}b) the radiation pattern is quite similar
to the one obtained for single gluon exchange.  Hard
emitted gluons have too short a wavelength to see the screening of the
color charge of the harder exchanged gluon by the second, typically very
soft, exchanged gluon. The Low-Nussinov pomeron thus reveals itself as an
extended object. % with a typical size of the order of $1/\Lambda_{\rm QCD}$.
Hard gluon emission is able to resolve the internal color structure of the
Pomeron.

As the transverse momentum of the emitted gluon is decreased, a qualitative
transition occurs, as is apparent by comparing the $p_{Tg}=2$~GeV and 15~GeV
cases in Fig.~\ref{figthree}. The gluon radiation has too long a wavelength to
resolve the internal color structure and hence the pomeron appears as a color
singlet object. As a result the
emission of a soft gluon ($p_{Tg}<<p_{Tq}$) follows a pattern very similar
to the one observed for $t$-channel photon exchange. This pattern is
expected to lead to the formation of rapidity gaps. Since the overall gluon
emission rate is dominated by the soft region, one concludes that two gluon
color singlet exchange in dijet events may indeed lead to the formation of
rapidity gap events as observed at the Tevatron~\cite{brandt}.

%\acknowledgements
{\bf Acknowledgements}
This research was supported in part by the University of Wisconsin Research
Committee with funds granted by the Wisconsin Alumni Research Foundation and
by the U.~S.~Department of Energy under contract No.~DE-AC02-76ER00881.

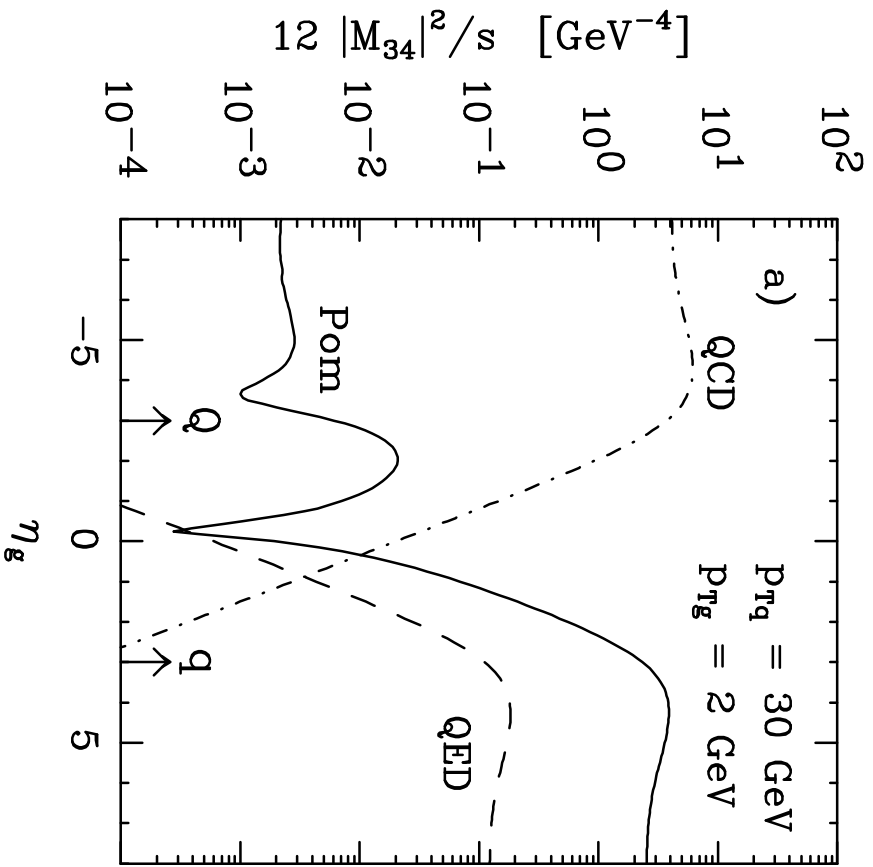
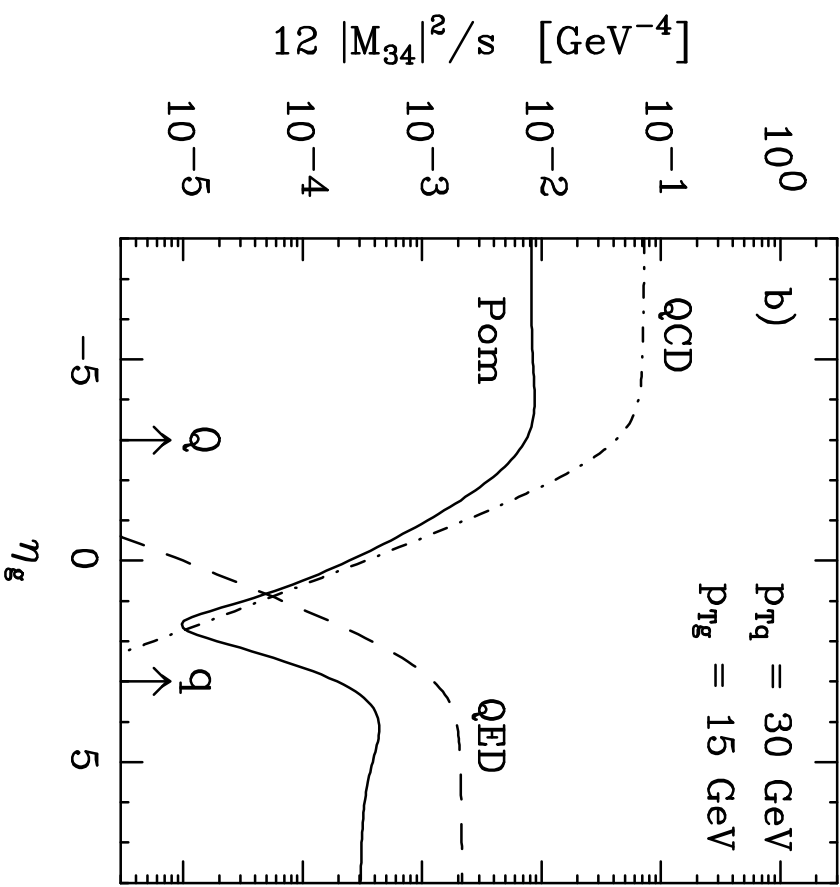
\end{document}